# Effect of Random Parameter Switching on Commensurate Fractional Order Chaotic Systems

Saptarshi Das[1,2*], Indranil Pan[3], and Shantanu Das[4]

*1. Department of Power Engineering, Jadavpur University, Salt Lake Campus, LB-8, Sector 3, Kolkata-700098, India.*

*2. Department of Physics, University of Cambridge, JJ Thomson Avenue, Cambridge CB3 0HE, United Kingdom.*

*3. Department of Earth Science and Engineering, Imperial College London, Exhibition Road, SW7 2AZ, United Kingdom.*

*4. Reactor Control Division, Bhabha Atomic Research Centre, Mumbai-400085, India.*

**Email:**

saptarshi@pe.jusl.ac.in, sd731@cam.ac.uk (S.Das*)

i.pan11@imperial.ac.uk, indranil.jj@student.iitd.ac.in (I. Pan)

shantanu@barc.gov.in (Sh. Das)

**Corresponding author's phone number:** +44(0)7448572598

## Abstract

The paper explores the effect of random parameter switching in a fractional order (FO) unified chaotic system which captures the dynamics of three popular sub-classes of chaotic systems i.e. Lorenz, Lu and Chen's family of attractors. The disappearance of chaos in such systems which rapidly switch from one family to the other has been investigated here for the commensurate FO scenario. Our simulation study show that a noise-like random variation in the key parameter of the unified chaotic system along with a gradual decrease in the commensurate FO is capable of suppressing the chaotic fluctuations much earlier than that with the fixed parameter one. The chaotic time series produced by such random parameter switching in nonlinear dynamical systems have been characterized using the largest Lyapunov exponent (LLE) and Shannon entropy. The effect of choosing different simulation techniques for random parameter FO switched chaotic systems have also been explored through two frequency domain and three time domain methods. Such a noise-like random switching mechanism could be useful for stabilization and control of chaotic oscillation in many real-world applications.

# 1. Introduction

Fractional calculus has given impetus to the study of dynamical systems which give rise to chaos and are applicable in diverse disciplines like physics, biology, economics etc. [1]. Fractional order nonlinear systems governed by fractional order differential equations (FODEs) can exhibit chaotic phenomena even for an overall order less than three. This contradicts with the classical theory of integer order (IO) nonlinear dynamical systems that in a continuous time nonlinear system, chaos can only manifest with at least an overall order of three [2]. This is due to the fact that each FO differ-integral operators are actually infinite dimensional systems which manipulate the underlying dynamics of each state with a very high order linear filtering (under a chosen rational approximation technique). However such investigations are not without their pitfalls, as higher order rational approximations of such FO operators have also led to spurious notions of chaos, termed as fake chaos [1]. Classical way of detecting chaos is carried out by the study of Lyapunov exponents from the system's parametric structure or from the observation of states in the case of unavailability of system's governing equations [3]. Chaos can be detected when the system has at least one positive Lyapunov exponent. If a system with four state equations has two positive Lyapunov exponents the system is known as a hyper-chaotic system.

After the exhaustive study of fractional calculus in linear control theory for the last decade [4], [5], it has pervaded into the nonlinear dynamical systems theory as well [1]. For example many classical chaotic systems have been extended with its analogous FO versions e.g. Chua system [2], Lorenz system [6], Chen system [7], Lu system [8], unified chaotic system [9], Rossler system [7], Liu system [10], Duffing oscillator [11], Van-der Pol oscillator [12], Lotka-Volterra model [13], financial system [14], Newton–Leipnik system [15], Volta's system [16], Arnedo system [17], Genesio–Tesi system [18], neuron network system [19], memristor based system [20], micro-electromechanical system [21], multi-scroll chaotic attractors [22], [23] and multi-wing chaotic attractors [24] and FO system without equilibrium points [25] etc. Similar traces of FO hyper-chaos like hyper-chaotic-Rossler system [7], Lorenz system [26], Chen system [27] and Four-wing attractor [28] with more than one positive Lyapunov exponents, have also been found. Beside the extension of FO versions of many well-researched chaotic systems mentioned earlier, FODEs are commonly observed to model various natural systems where the temporal dynamics has got a long-memory behavior like $1/f^\alpha$ noise, heavy-tailed distribution, long-range dependency in the auto-correlation, power-law decay in excitation-relaxation systems (like electrical circuits, spring-mass-damper system, chemical reactions etc.) instead of an exponential envelope [4], [29]. Fractional dynamics is more evident in time domain representation of naturally occurring oscillatory systems (like viscoelasticity, fluid flow etc. [30], [31]) where the notion of damping is different with a decaying power-law envelope and Mittag-Leffler type oscillations [32], instead of sinusoidal oscillations within an exponentially decaying envelope.

The focus of the present study is to first observe chaos in the unified chaotic system family that encompasses the dynamics of three popular classes of chaotic systems (i.e. Lorenz, Lu and Chen) for different ranges of a single key parameter ($δ$). Next, a random noise like fast switching in the parameter $δ$ is proposed, such that the system's behavior continuously jumps between these three families of chaotic attractors. In addition, the effect of decreasing the commensurate FOs have been elucidated for the random parameter system to discriminate the new phenomena in comparison with the traditional fixed parameter FO chaotic systems mentioned above.

The rest of the paper is organized as follows. Section 2 briefly describes the unified FO chaotic system family, different time/frequency domain simulation techniques, error analysis and the effect of different numerical integration solvers. Section 3 shows the simulations for the random noise like fast switching in the unified chaotic system key parameter $\delta$ and the suppression of chaos particularly at low values of the commensurate FO. Section 4 shows the characterization of different random parameter chaotic systems using two system diagnostics – LLE and Shannon entropy. Section 5 sheds light on the novelty of the observation and the analysis of existence of chaos in the proposed random parameter switched Unified system. The paper ends with the conclusions in section 6, followed by the references.

## 2. Fractional Order Unified Chaotic Systems

### 2.1. Unified Chaotic System Family

The IO unified chaotic system (1) models the classical Lorenz-Lu-Chen family of chaotic attractors with a common template and a single key parameter $\delta$ that determines which family the chaotic system will belong to.

$$\begin{aligned}
\dot{x} &= (25\delta + 10)(y - x) \\
\dot{y} &= (28 - 35\delta)x - xz + (29\delta - 1)y \\
\dot{z} &= xy - (\delta + 8)z/3
\end{aligned} \quad (1)$$

where, $\delta \in [0,1]$. The unified chaotic system behaves as the family of either Lorenz, Lu or Chen system within the following ranges of parameter $\delta$ respectively [33].

$$\begin{aligned}
\delta \in [0, 0.8) &\Rightarrow \text{Lorenz family} \\
\delta = 0.8 &\Rightarrow \text{Lu system} \\
\delta \in (0.8, 1] &\Rightarrow \text{Chen family}
\end{aligned} \quad (2)$$

Recent studies [9] have developed analogous FO chaotic system by replacing the IO derivative terms by equivalent fractional derivatives, as shown in (3) with the commensurate FO $\alpha \in (0,1]$ which still exhibits chaotic nature depending on the value of the order of the FO nonlinear system $\alpha$.

$$\begin{aligned}
\frac{d^{\alpha}x}{dt^{\alpha}} &= (25\delta + 10)(y - x) = f_1(x, y, z) \\
\frac{d^{\alpha}y}{dt^{\alpha}} &= (28 - 35\delta)x - xz + (29\delta - 1)y = f_2(x, y, z) \\
\frac{d^{\alpha}z}{dt^{\alpha}} &= xy - (\delta + 8)z/3 = f_3(x, y, z)
\end{aligned} \quad (3)$$

### 2.2. FO Chaotic System Simulation Techniques Using Frequency and Time Domain Methods

It has been shown in many literatures that FO nonlinear dynamical systems exhibit chaotic behaviour even when the commensurate order is less than one i.e. the overall system order being lesser than three which is impossible in the case of classical IO chaotic systems [2]. Numerically solving such FO chaotic systems is a bit challenging as compared to the IO chaotic systems. The Adams-Bashforth-Moulton predictor-corrector method is widely adopted in various literatures to avoid appearance of fake chaos or loss of chaos due to truncation or rational approximation by frequency domain methods of realizing fractional differ-integrals. But for noisy FO chaotic systems and fast randomly switched parameter systems, extension of such time domain numerical solvers similar to Adams-Bashforth-Moulton method has not been explored yet.

Petras [1] has suggested that as an alternative way, the FO differential equations may be converted to a set of integral equations with fractional derivative less than unity as shown in equation (4) for efficiently solving such equations. In this way, the fractional derivatives can be rationalized using Oustaloup's/modified Oustaloup's or any other frequency domain approximation methods [5].

$$\begin{aligned} x(t) &= {}_0D_t^{1-\alpha} \int_0^t \left[ (25\delta + 10)(y-x) \right] dt \\ y(t) &= {}_0D_t^{1-\alpha} \int_0^t \left[ (28 - 35\delta)x - xz + (29\delta - 1)y \right] dt \\ z(t) &= {}_0D_t^{1-\alpha} \int_0^t \left[ xy - (\delta + 8)z/3 \right] dt \end{aligned} \quad (4)$$

where, ${}_0D_t^{1-\alpha}$ is Riemann-Liouville fractional differentiation operator.

In the present study, we have primarily used the modified Oustaloup's recursive approximation for each fractional derivative with a 30$^{th}$ order approximation within a chosen frequency range of $[\omega_b, \omega_h] = [10^{-4}, 10^4]$ Hz using the FOMCON Toolbox [34] in the Matlab/Simulink environment to simulate such FO chaotic systems as suggested in Petras [1]. The modified Oustaloup's approximation converts the fractional differ-integrals in Riemann-Liouville definition (5) in equivalent frequency domain Infinite Impulse Response (IIR) filters given by (6).

$$\begin{aligned} {}_aD_t^\alpha f(t) &= \frac{1}{\Gamma(m-\alpha)} \frac{d^m}{dt^m} \int_a^t \frac{f(\tau)}{(t-\tau)^{\alpha-m+1}} d\tau, (m-1) < \alpha < m, \alpha \in \mathbb{R}^+ \\ \mathcal{L}\{{}_0D_t^\alpha f(t)\} &= s^\alpha F(s) - \sum_{k=0}^{m-1} s^k \left[ {}_0D_t^{\alpha-k-1} f(t) \right]_{t=0} \end{aligned} \quad (5)$$

Here, $s$ and $t$ denotes the Laplace and time domains respectively. For zero initial conditions the Laplace transform of Riemann-Liouville fractional derivatives reduces to $\mathcal{L}\{{}_0D_t^\alpha f(t)\} = s^\alpha F(s)$ [1] [5].

$$s^\alpha \approx \left( \frac{d_f \omega_h}{b_f} \right)^\alpha \left( \frac{d_f s^2 + b_f \omega_h s}{d_f (1-\alpha) s^2 + b_f \omega_h s + d_f \alpha} \right) \prod_{k=-N}^{N} \frac{s + \omega_k'}{s + \omega_k} \quad (6)$$

$$\omega_k = \omega_b \omega_u^{\left(\frac{2k-1+\alpha}{N}\right)}, \omega_k' = \omega_b \omega_u^{\left(\frac{2k-1-\alpha}{N}\right)}, \omega_u = \sqrt{\omega_h/\omega_b}$$

The choice of refined Oustaloup's filter parameters ($b_f$ and $d_f$) in (6) has been described in [35] in an optimization based framework where it has been shown that the constant phase performance remains almost consistent with a fixed $b_f/d_f$ ratio, whereas the original proposition for modified Oustaloup's filter suggested a fixed parameter of $b_f = 10, d_f = 9$. The particular advantage of using modified Oustaloup's approximation instead of the original Oustaloup's approximation (implemented in another popular Matlab based Toolbox for FO systems, called Ninteger [36]) is that the former gives lesser ripple in the frequency domain phase error surface. Therefore it maintains a constant phase for a wider frequency range and gives good approximation even near the chosen boundaries of lower/upper frequency [35], [5]. The original Oustaloup's filter's poles, zeros and gain are recursively calculated as:

$$s^\alpha \approx K \prod_{k=-N}^{N} \frac{s+\omega'_k}{s+\omega_k}, \omega_k = \omega_b \omega_u^{\left(\frac{2k-1+\alpha}{N}\right)}, \omega'_k = \omega_b \omega_u^{\left(\frac{2k-1-\alpha}{N}\right)}, K = \omega_h^\alpha \qquad (7)$$

For the present study in both the frequency domain approximation techniques for FO differ-integrators, the nonlinear FO ordinary differential equations have been integrated with the fourth order Runge-Kutta method with fixed step size of 0.001 sec.

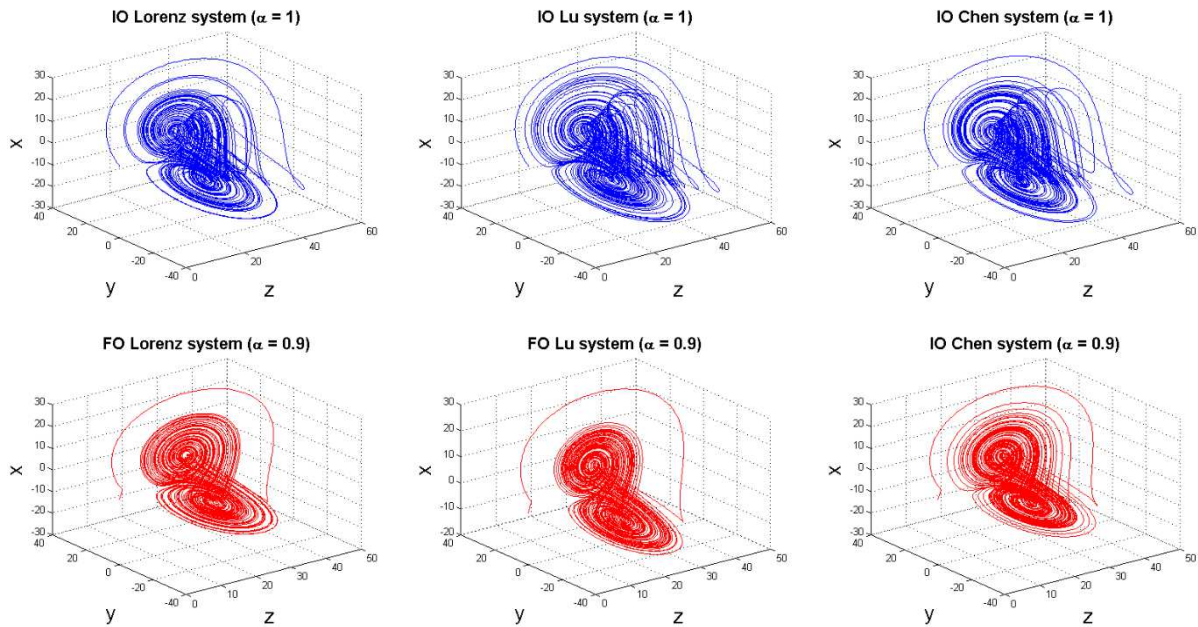

*Figure 1: Phase portraits of IO (α = 1) and commensurate FO (α = 0.9) chaotic Lorenz, Lu and Chen system (ᾱ = 0.9, 0.8, 0.7 respectively)*

The frequency domain rational approximation methods sometimes may be less accurate for FO differential equations, especially for the investigation of chaos. Tavazoei and Haeri [37], [38] have shown that chaos may disappear in a FO system due to frequency domain rational approximation whereas a non-chaotic nonlinear FO system may appear to be chaotic due to such rational approximation. This is because the frequency domain rational approximation techniques modifies the nature, number and stability of the fixed points. Therefore, time domain methods for fractional differential equations should also be tested alongside to verify that the nature of phase space response is consistent between the frequency and time domain computation of the FODEs. The most popular time domain method for single and multi-term FODEs is known as the Adams-Bashforth-Moulton PECE (predict, evaluate, correct, evaluate) method [39], [40]. Garrappa came up with a new implementation of the PECE algorithm using Fast Fourier Transform (FFT) for the convolutions which reduces computational cost of the Volterra convolution equation. The stability of Garrappa's implementation of PECE method has been studied in [41] and its Matlab based implementation is known as *fde12*. There is also another class of time domain method known as the implicit fractional linear multistep methods (FLMMs) of the second order. The FLMM has three subclasses *viz.* Trapezoidal method, Newton-Gregory (NG) formula, and backward differentiation formula (BDF) which are implemented by Garrappa in Matlab as *flmm2* algorithms and the detailed mathematical treatments of PECE and FLMM can be found in [42], [42], [43] respectively. Here, we focus on exploring the difference in the resulting state trajectories using these different time domain and frequency domain simulation techniques.

The PECE method needs only the functional form of the nonlinear state equations, whereas the FLMM methods additionally needs the Jacobian information which is given by (8) for the FO Unified system (3). Detailed error analysis of Jacobian based predictor corrector method has been reported in [44].

$$J_f = \begin{bmatrix} \partial f_1/\partial x & \partial f_1/\partial y & \partial f_1/\partial z \\ \partial f_2/\partial x & \partial f_2/\partial y & \partial f_2/\partial z \\ \partial f_3/\partial x & \partial f_3/\partial y & \partial f_3/\partial z \end{bmatrix} = \begin{bmatrix} -(25\delta+10) & (25\delta+10) & 0 \\ (28-35\delta)-z & (29\delta-1) & -x \\ y & x & -(\delta+8)/3 \end{bmatrix} \quad (8)$$

In order to use the existing three time domain FLMM solvers for the simulation of random parameter FO chaotic systems (as explored in subsequent sections), the same random choice of the key parameter is declared as a global variable and then passed in both the functions containing the system structure (3) and the Jacobian (8) simultaneously. Choosing different random key parameter in the system structure and Jacobian at one time step do not yield convergence of the FLMM methods. Whereas, in the PECE method, the Jacobian information is not required. Therefore the key parameter $\delta$ can directly be randomised within the function containing the system structure and not to be passed from outside as a global variable like in the three FLMM methods.

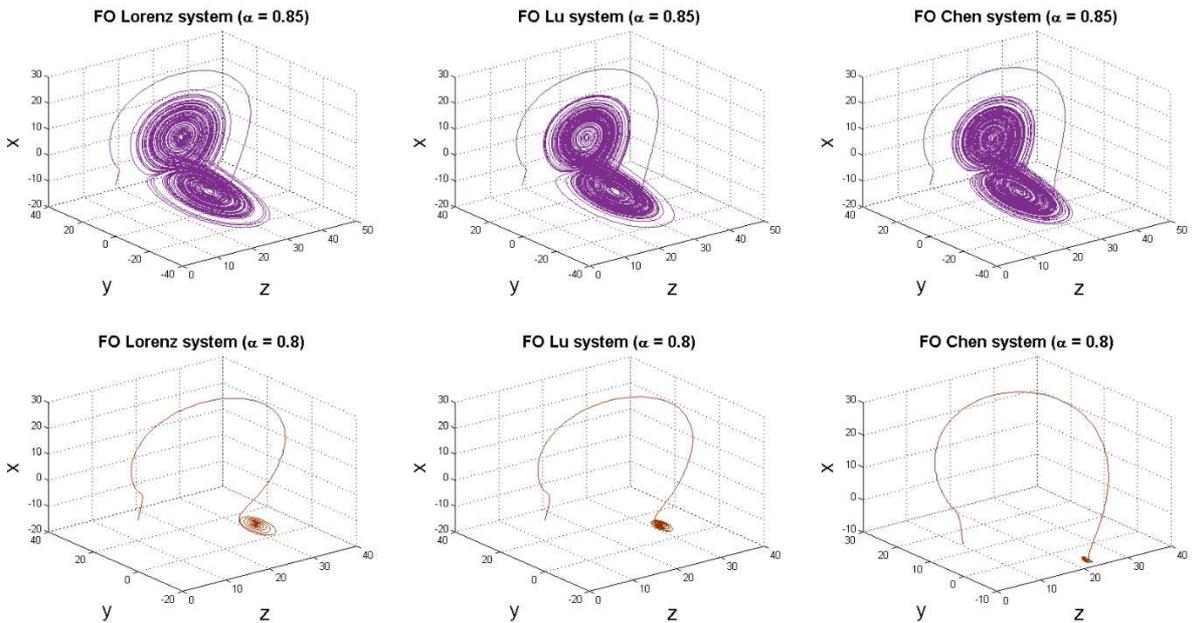

*Figure 2: Disappearance of chaos in FO unified chaotic system family for commensurate order* α ≤ *0.8*

As mentioned earlier, the unified chaotic system behaves as the family of Lorenz, Lu and Chen systems for different ranges of the key parameter δ as given in (2). The phase portraits of the three IO chaotic systems (Lorenz-Lu-Chen) and that of their commensurate FO counterparts (with α = 0.9) are shown in Figure 1 with an initial condition of $(x_0, y_0, z_0) = (1, 3, 2)$ using the frequency domain refined Oustaloup's approximation method. It is found that decrease in the commensurate FO of the unified system still exhibits chaotic phase portraits till α = 0.85. For further decrease in commensurate FO up to α = 0.8, the phase space trajectory stabilizes in one of the stable equilibrium points as shown in Figure 2. Therefore, for FO systems, chaos can appear even in a system having an overall order $0.85 \times 3 = 2.55$. Tavazoei and Haeri in [45] have shown the bounds of observing chaos in commensurate and incommensurate FO unified chaotic system and in general, the gradual decrease in FO causes disappearance of chaos.

### 2.3. Error Analysis of Time and Frequency Domain Methods for the Numerical Simulation of Chaotic FODEs

The error analysis for numerical simulation of chaotic FODEs have been carried out in two different ways –

- Error incurred in the frequency domain approximation (while maintaining a constant phase response within a chosen frequency band) for the FO elements

- Error (rather difference) between the state trajectories using different numerical methods to integrate the chaotic FODEs

To explain the first point - FO differ-integrators are known as constant phase elements (CPEs) and different rational approximations try to maintain a nearly constant phase response by placing interlaced poles and zeros, over a chosen frequency band. However the error between the original fractional differ-integrator and their rational approximation are affected by the choice of the method, frequency bandwidth and order of realization. Figure 3 shows the error between the ideal and approximated phase responses for Oustaloup's and modified Oustaloup's method for different FOs (α). It is evident from Figure 3 that the refined Oustaloup's filter produces less error in the approximation of a smooth phase response especially near the boundaries $(\omega_b, \omega_h)$. Here, the phase error is given by (9) as the difference between ideal and approximated response for a FO differ-integrator.

$$\Delta \phi = \phi_{\text{approx}} - (\alpha \pi / 2) \qquad (9)$$

We have also shown the variation in the phase error with the order of approximation in both the Oustaloup's and refined Oustaloup's method in Figure 4. It is evident that in both the methods for low order of rational approximation ($N < 10$), there is some ripple in the phase response and after that there is a small variation in accuracy as also found previously for several other rational approximation techniques like Carlson, continued fraction expansion (CFE) in continuous and discrete time etc. [35], [46]–[48]. However, as reported in Petras [1], we chose a 30[th] order approximation for both the Oustaloup's and refined Oustaloup's approximation of FO operators in the FODEs.

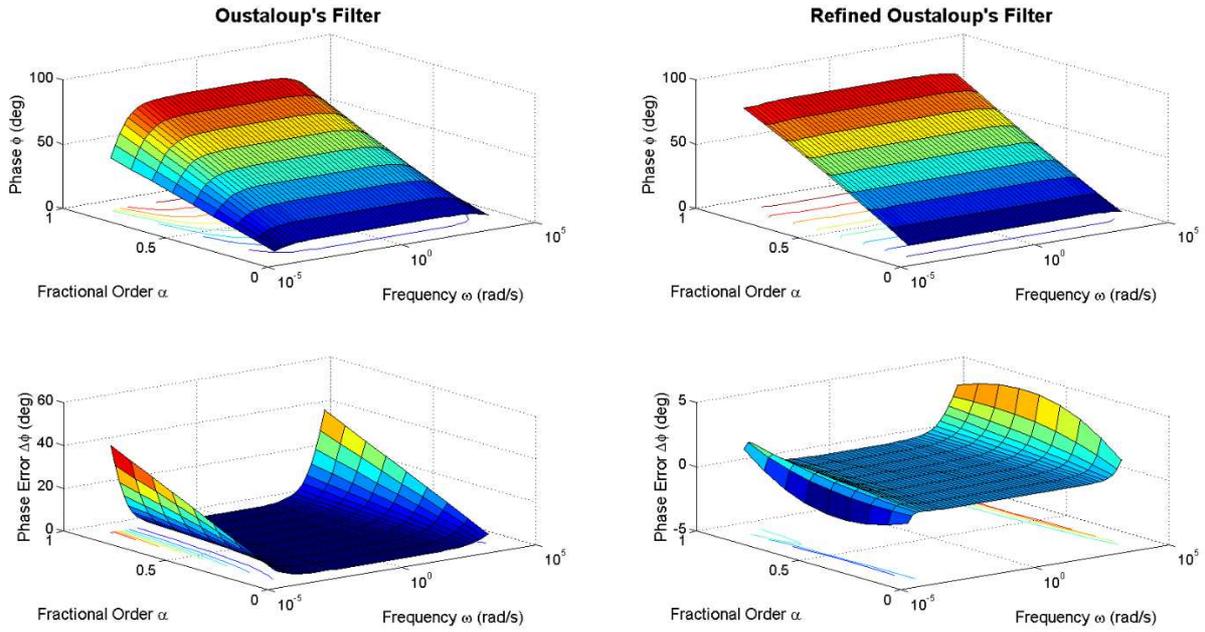

*Figure 3: Phase ripple in Oustaloup's and refined Oustaloup's method with variation in FO (α).*

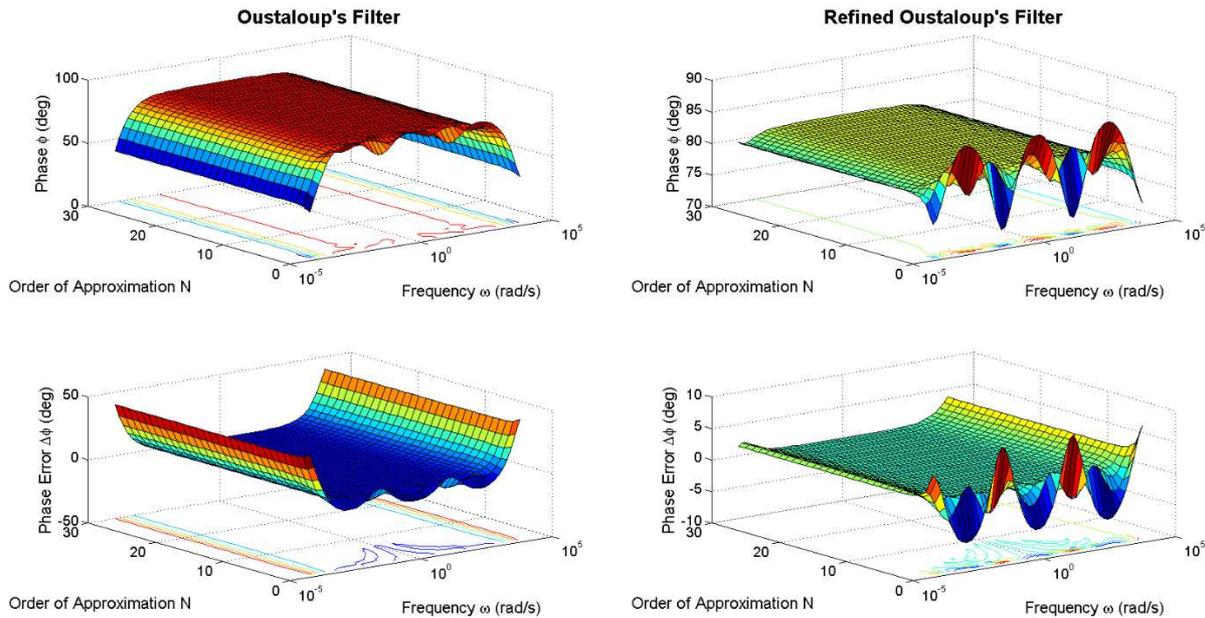

*Figure 4: Phase ripple in Oustaloup's and refined Oustaloup's method with variation in order of approximation (N).*

To explain the second point, we have also shown the difference between the most consistent time domain method – FLMM BDF (backward difference formula) and the other three variants of time domain methods (PECE, FLMM Trapezoidal, FLMM Newton-Gregory) along with two frequency domain methods (Oustaloup's and refined Oustaloup's method). Consistency of time domain methods for FODEs have already been discussed in Garrappa [43]. Since the frequency domain approximation methods try to maintain a constant phase response over the desired frequency range, the phase ripple with the change in approximation technique might have some impact on the accuracy of simulation for chaotic FODEs. However, here the objective is to find the consistency of simulations between frequency and time domain numerical methods for chaotic FODEs.

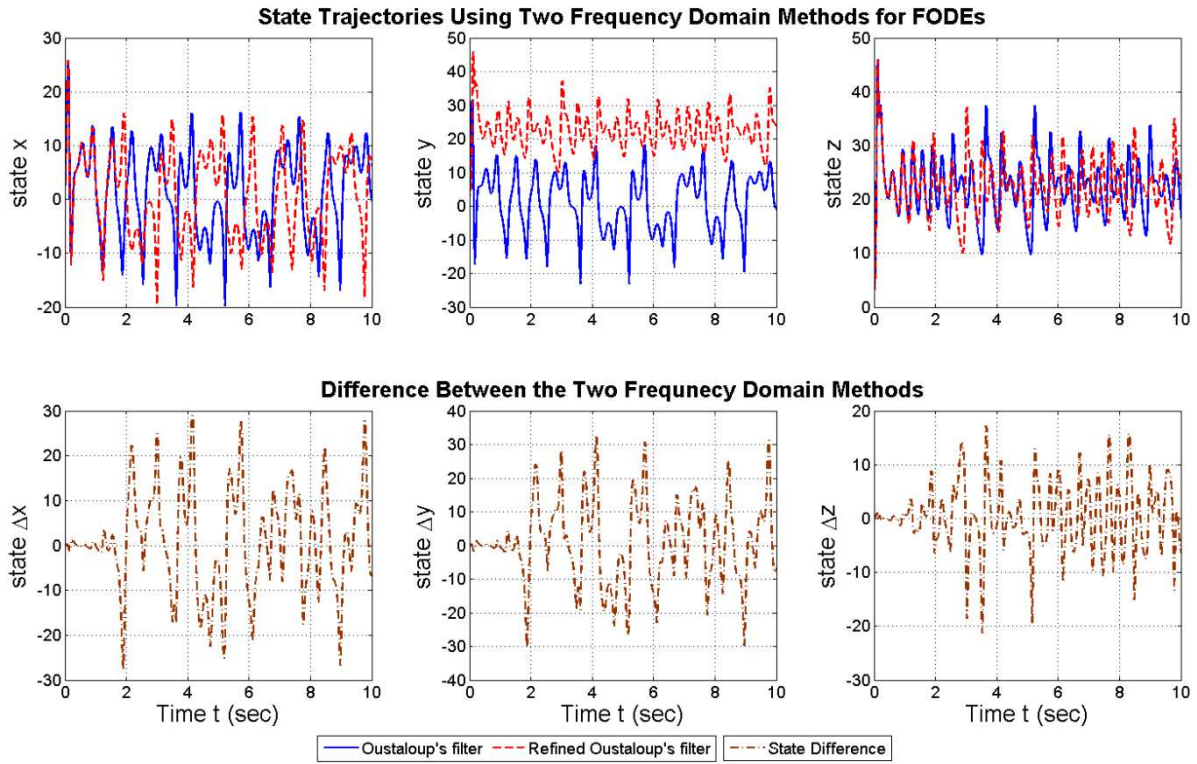

*Figure 5: State trajectories and their difference between two frequency domain methods for numerical integration of chaotic FO unified system.*

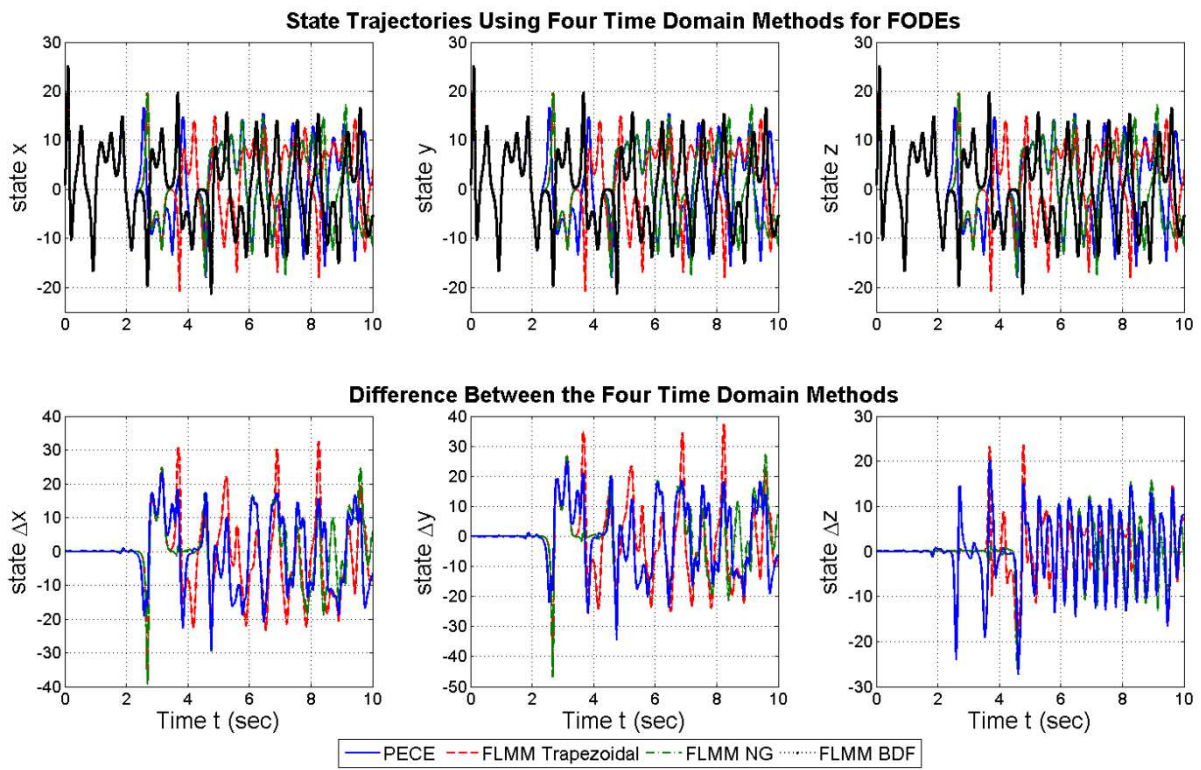

*Figure 6: State trajectories and their difference amongst four time domain methods for numerical integration of chaotic FO unified system.*

The difference between the state trajectories are shown in Figure 5 for the two frequency domain methods and in Figure 6 for the three time domain methods. This helps in understanding the difference in the temporal evolution of the state variables while using different techniques for FODEs. All of the five methods use some sort of approximation either in time or frequency domain to evaluate the state trajectories, no ground truth (base case) of the state variables can be ascertained to compute the error incurred in the evolution of the states. Therefore we have provided the comparative plots between these time and frequency domain techniques for a fixed parameter FO unified chaotic system. The robustness of these methods are judged for the random parameter case yielding minimum variance in estimating the LLE and entropy, in the coming sections. However for comparing the four time domain techniques, the FLMM BDF method has been considered as the benchmark solution for the FO chaotic system as suggested in [43] and rest of the three system's response have been compared with it for the time domain error analysis of the three state variables. The errors have been quantified using the 1-norm, 2-norm and ∞-norm difference between the state trajectories obtained with FLMM BDF vs. rest of the methods in Table 1. Considering the error signal being $e_i[n] = X_i^{FLMM-BDF}[n] - X_i^{other}[n], i \in \{x, y, z\}$ over a finite time length (here 5 sec and thus sample size $n = 5 \times 10^4$) these three norm difference are calculated as:

$$\Delta L_1 = \sqrt{\sum_n e[n]}, \Delta L_2 = \sqrt{\sum_n e^2[n]}, \Delta L_\infty = \max(|e[n]|) \quad (10)$$

The best results (in terms of $L_p$ norm difference) between the FLMM BDF and other time/frequency domain methods are highlighted in bold-italics. It is seen that the PECE is better most of the times and is able to maintain close enough state trajectories to that of the FLMM BDF.

Table 1: $L_p$ Norm difference ($p = 1, 2, \infty$) of the three state errors with respect to the FLMM BDF time domain method for time length of 50 sec

| State | Error Norm | FLMM Trapezoidal | FLMM NG | PECE | Oustaloup | Refined Oustaloup |
|---|---|---|---|---|---|---|
|   | $\Delta L_1$ | 460655.21 | *__438143.61__* | 439773.98 | 438799.51 | 457390.97 |
|   | $\Delta L_2$ | 2562.89 | 2441.84 | *__2386.24__* | 2439.07 | 2502.79 |
| x | $\Delta L_\infty$ | 39.39 | 39.42 | *__32.20__* | 34.21 | 34.54 |
|   | $\Delta L_1$ | 490033.46 | *__469082.11__* | 471718.70 | 471085.30 | 475514.32 |
|   | $\Delta L_2$ | 2741.20 | 2624.85 | *__2564.67__* | 2627.15 | 2597.90 |
| y | $\Delta L_\infty$ | 46.65 | 46.70 | *__37.17__* | 39.23 | 38.18 |
|   | $\Delta L_1$ | 460655.21 | *__438143.61__* | 439773.98 | 438799.51 | 457390.97 |
|   | $\Delta L_2$ | *__1418.53__* | 1497.70 | 1627.68 | 1457.53 | 5243.90 |
| z | $\Delta L_\infty$ | *__23.62__* | 26.17 | 27.22 | 23.69 | 50.45 |

## *2.4. Effect of Numerical Integration Methods on the Frequency Domain Simulation of chaotic FODEs*

For the numerical integration of IO fixed parameter chaotic systems, often the variable and adaptive step size solvers are used. However, it is evident from the simulated phase portraits in Figure 1 and Figure 2 as well as the calculated Lyapunov exponents and entropy values (in the next sections) that integrating the integer and fractional order Unified chaotic system is sufficient to investigate the chaotic behaviour of the state variables. Also, the effect of random parameter perturbation can be more visible and quantifiable using various system diagnostic measures with a regular time step method instead of a variable step one. Also, the calculation of Lyapunov exponent from the noise-like time series of the state trajectories becomes an extremely difficult task if the state is irregularly sampled due to applying a variable step-size integration method. However, to address the question whether our method for numerically integrating the FO chaotic system with fixed step-size of 0.001 sec is robust enough, we here show exhaustive simulation study each for a fixed step-size of 0.0001 and 0.00001 sec respectively. We also quantitatively show that this granularity of simulation does not appreciably change the results in anyway, but introduces a huge computational burden. The computational time for varying all the three step size methods of numerical integration are compared in Table 2, when 50 sec of simulations have been run in Matlab/Simulink environment on a 64 bit Windows 7 desktop PC with Intel I5, 3.3 GHz processor and 16 GB of RAM. State trajectories of the fixed parameter FO chaotic system (with $\delta = 0.8$, $\alpha = 0.9$ and refined Oustaloup's rational approximation method) have also been explored in Figure 7, using different fixed step ODE solvers with order of accuracy varying between one and five.

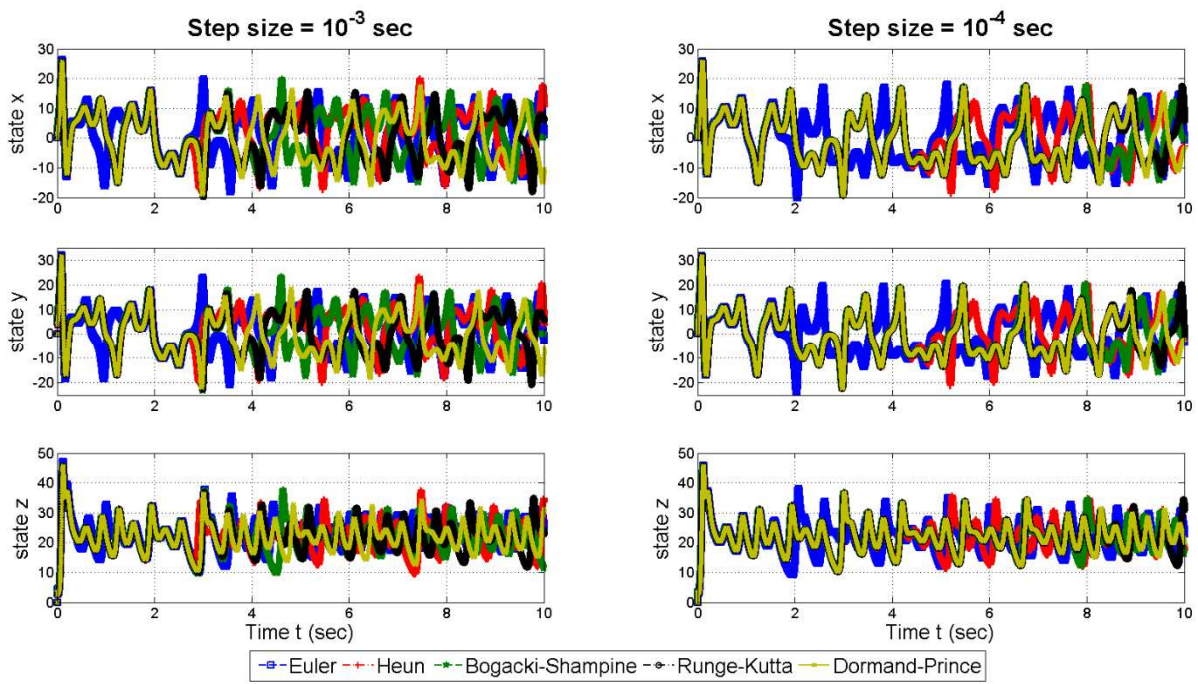

*Figure 7: Difference in the evolution of state trajectories with change in numerical integration solver for fixed parameter FO chaotic system with $\delta = 0.8$, $\alpha = 0.9$*

Now, we explore the effect of varying the fixed step-size explicit solver to integrate the chaotic FODEs yielding a first (Euler's method), second (Heun's method), third (Bogacki-Shampine method), fourth (Runge-Kutta method) and fifth (Dormand-Prince method) order of accuracy along with a comparison of their runtime reported in Table 2. In [49], it is shown that the RK4 (fourth order Runge Kutta) method can be considered as a robust solver for a wide variety of nonlinear stiff ODEs. Therefore in the frequency domain approximation methods, we have adopted the Runge-Kutta method as a trade-off between accuracy and simulation run-time, since higher order methods are more computationally intensive as shown in Table 2. Figure 7 also shows that starting from the same initial condition but with a finer step size, the state trajectories deviate later between the five different choices of numerical integration solvers with an order of accuracy varying between one and five. It also shows that the evolution of the states gradually starts departing from the initial condition while using different solvers to numerically integrate the ODEs, but the information content or entropy estimates for the random parameter chaotic system is not greatly affected by the change of solvers as shown in Table 2. Also in order to make a fair comparison with the time domain methods (PECE and FLMM) we have restricted our study to fixed step size solvers only, as these time domain methods for FODEs use fixed step size too. In addition, the system diagnostics e.g. the LLE and Shannon entropy calculation becomes extremely complicated if the continuous states are sampled with irregular interval, as a result of variable step size integration.

Table 2: Effect of changing the solver for numerical integration on the computational time (for simulation time of 50 sec) based on refined Oustaloup's approximation for fixed parameter unified system with α = 0.9, δ = 0.8

| Solver | Order of Accuracy | step-size (sec) | Computational Time (sec) | Shannon Entropy $H_x$ (for state x) |
|---|---|---|---|---|
| Euler | 1 | $10^{-3}$ | 4.27 | -25123.46 |
|  |  | $10^{-4}$ | 11.83 | -241890.40 |
|  |  | $10^{-5}$ | 425.59 | -2518606.68 |
| Heun | 2 | $10^{-3}$ | 4.26 | -24336.90 |
|  |  | $10^{-4}$ | 15.65 | -248293.92 |
|  |  | $10^{-5}$ | 963.98 | -2433222.74 |
| Bogacki-Shampine | 3 | $10^{-3}$ | 4.94 | -26430.05 |
|  |  | $10^{-4}$ | 20.15 | -240143.35 |
|  |  | $10^{-5}$ | 674.66 | -2473862.78 |
| Runge-Kutta | 4 | $10^{-3}$ | 5.23 | -24181.12 |
|  |  | $10^{-4}$ | 24.62 | -235823.72 |
|  |  | $10^{-5}$ | 549.38 | -2417141.51 |
| Dormand-Prince | 5 | $10^{-3}$ | 6.24 | -24651.35 |
|  |  | $10^{-4}$ | 37.25 | -238121.71 |
|  |  | $10^{-5}$ | 765.32 | -2479454.18 |

## 3. Random Noise Like Fast Switching in the Unified Chaotic System Parameter

Here, we explore a new class of chaotic systems with random switching in its key parameter *δ*. The IO and FO unified chaotic system is chosen for the simulation study with a consideration of random switching in the key parameter $\delta \sim \mathcal{U}(0,1)$. Therefore, the key system parameter *δ* of the unified chaotic system is randomly drawn from a uniform distribution such that at each time instant it belongs to a particular system family according to (2) but at the next time instant it jumps to another chaotic system family. The small wiggles in the phase space trajectories even for the IO case is evident in Figure 8 due to random variation in the unified chaotic system key parameter *δ*, as opposed to the smooth phase space trajectories in IO unified chaotic systems in Figure 1. The aberrations in the phase space trajectories are also evident for lower value of the commensurate FO of *α* = 0.9 as shown in Figure 9. The change in the phase space trajectories for such random noise like fast parameter switching FO systems indicate towards a new type of chaotic behaviour, as compared to that shown in Figure 1 and Figure 2, for classical fixed parameter IO/FO chaotic systems. Systems exhibiting this kind of behaviour are henceforth called as IO/FO *random parameter switched chaotic systems* in the remainder of the paper.

It is also expected that for such a new class of random parameter FO switched chaotic systems the obtained state-time series would be totally different compared to the classical versions. Here, starting from the same initial condition $(x_0, y_0, z_0)$ and same initial parameter $\delta = 0.8$, the system's state variables evolve in a different way and diverges very quickly from each other due to the presence of random switching in the parameters of the chaotic system. Although the deterministic long term prediction property holds for the fixed parameter chaotic system if the underlying model is known but for the latter case due to the uncertainty introduced in the value of key system parameter (δ) due to noisy switching at each time instant, such prediction becomes impossible. As can be seen from the phase portraits in Figure 8-Figure 9, their shapes qualitatively match with the traditional cases with a constant key parameter in Figure 1-Figure 2, except the small wiggles on the phase space trajectories.

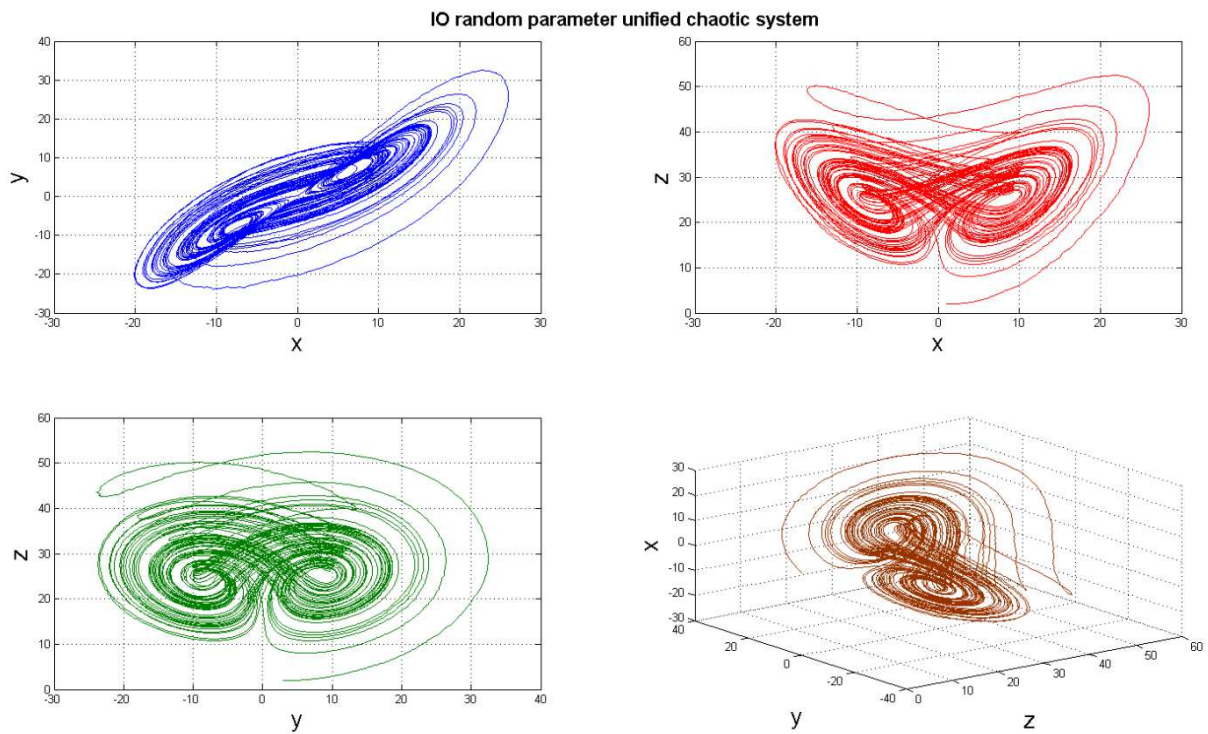

*Figure 8: Phase portrait of IO (α = 1) unified chaotic system with random variation in δ with wiggly trajectories.*

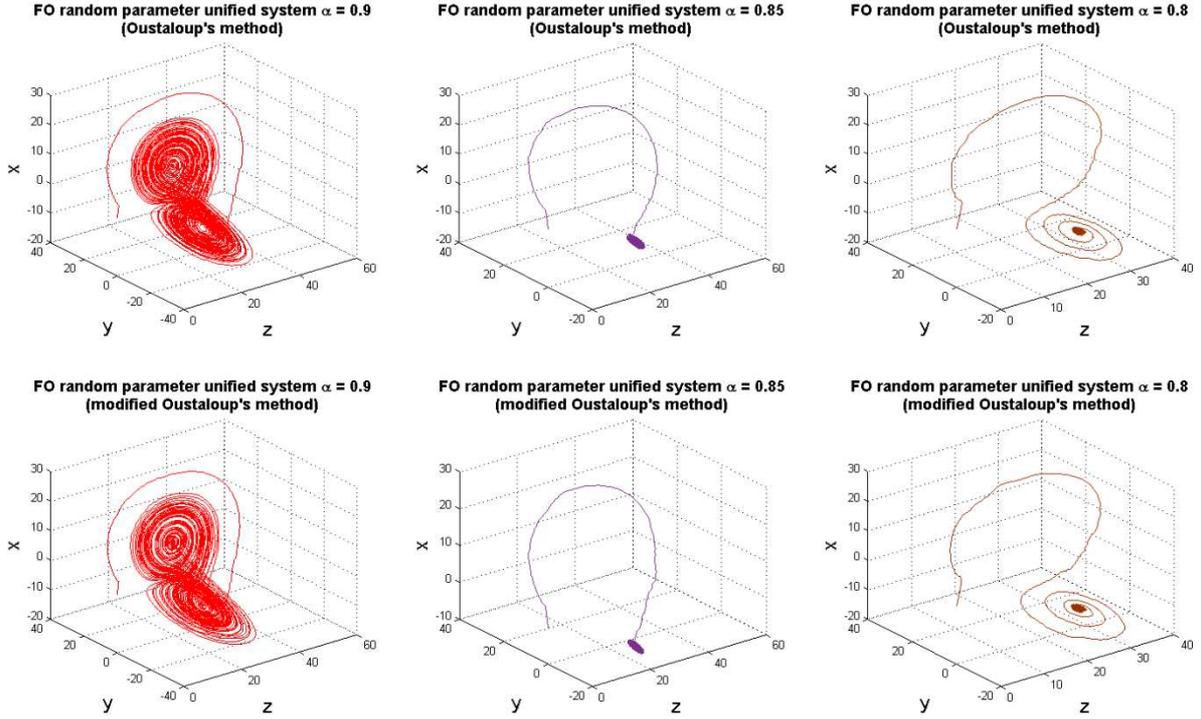

*Figure 9: Phase* portrait of FO unified chaotic system with random variation in δ with *α* = {0.9, 0.85, 0.8} with wiggly trajectories. Chaos disappears much early at *α* = 0.85, due to random key parameter (*δ*) switching whereas the fixed parameter unified system is still chaotic at *α* = 0.85.

The seed of the random number generator for the key parameter *δ* in different simulations were again randomly initialized 100 times for the same IO random parameter switched chaotic system so that the system evolves from the same initial condition $(x_0, y_0, z_0)$ but in different ways in successive simulation steps. The instantaneous values of the key parameter *δ* has been randomly changed by sampling from a uniform distribution as shown in Figure 10. It is also clear from Figure 10 that although in 100 different simulations, the state variables starts from the same initial value, they diverge very quickly due to the randomness introduced in the key parameter *δ*. This shows the unpredictability of the state even more than the traditional application of chaotic time series in the area of secure communication with a fixed initial conditions $(x_0, y_0, z_0)$ and model of the chaotic system (governed by affixed key parameter *δ* for encryption). Figure 10 also shows the histogram of the amplitudes of state variable and the key parameter.

Another interesting fact can be observed for such switched parameter FO chaotic systems in Figure 9. While gradually decreasing the commensurate FO, the chaos disappears much earlier than the fixed parameter FO chaotic systems. This typical phenomenon has been illustrated in Figure 9 where the phase space trajectories stabilizes in one of the equilibrium points for a commensurate FO of *α* = 0.85, whereas the fixed-parameter FO chaotic systems (Lorenz, Lu and Chen sub-classes) in Figure 2 still exhibit chaos at that low FO of *α* = 0.85. The early disappearance of chaos is found to be consistent with both the frequency domain methods – Oustaloup's and refined Oustaloup's method as shown in Figure 9 as well as the four time domain methods.

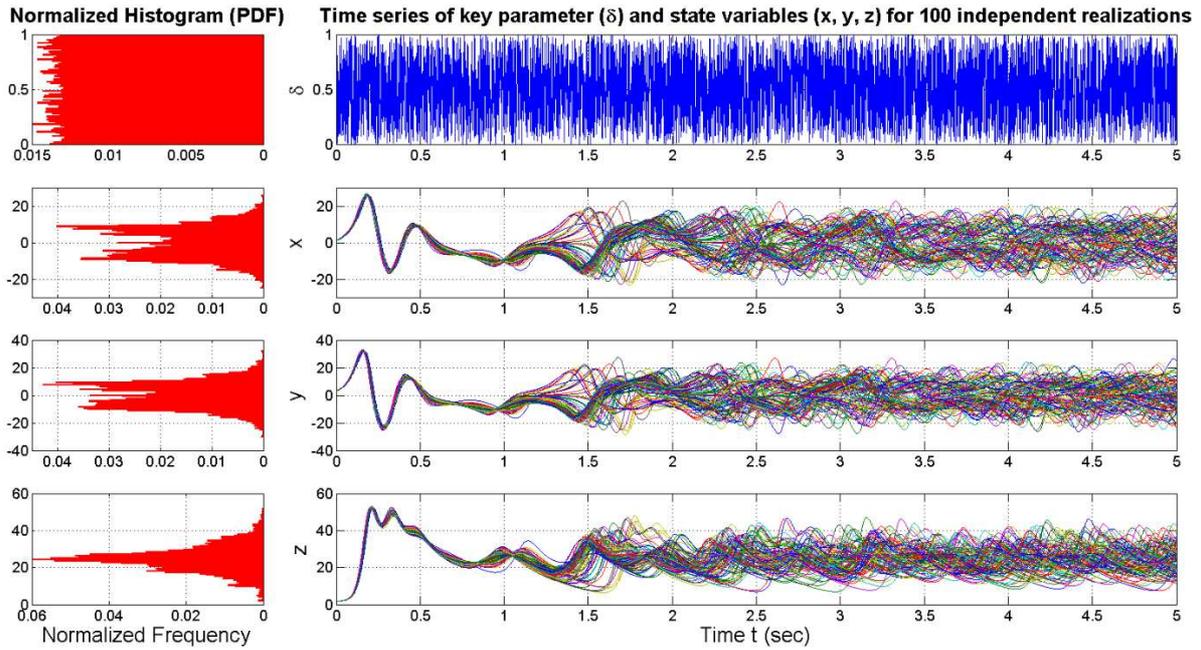

*Figure 10: Time series of the key parameter (δ) and the three state variables for the IO random parameter switched chaotic system with 100 independent realizations.*

This observation shows that such random parameter switching may be useful for control or suppression of chaotic phenomena in many naturally occurring chaotic high dimensional systems [50], where the task is to suppress chaotic oscillations or neutralize chaotic oscillations by noise [50], [51]. Previously, a similar concept has been introduced in [52], [53], [54], [55] known as noise induced chaos where the chaotic behaviour can be an effect of random parameter variation in a nonlinear system. The present approach is a bit different since the random variation in the key parameters ($\delta$) makes the system arbitrarily switch between the three families of attractors *viz.* Lorenz-Lu-Chen systems. The presence and disappearance of chaotic motion and randomness in the resulting time series in the previously explored fixed and random parameter chaotic systems are next verified using the numerical computation of Lyapunov exponents and Shannon entropy.

## 4. Diagnostics for the Time-Series of Random Parameter Switched Chaotic Systems Using Lyapunov Exponent and Shannon Entropy

Investigation of chaos is mostly done using its genuine signature known as the Lyapunov exponent. There has been significant amount of research in past to calculate the largest Lyapunov exponent (LLE) from a finite length time-series which is capable of discriminating the underlying dynamical behavior of a data due to random noise and chaotic motion. Lyapunov exponent greater than zero indicates the presence of chaotic fluctuation in a measured dataset. In most realistic cases, a measured time series is corrupted with white or colored noise which could result in spurious estimated values of LLE depending on the strength of noise added to the system. Investigation of the additive noise with chaotic time series with different signal to noise ratio (SNR) levels has been reported in Rosenstein *et al.* [56].

In the present paper, the problem is different since the dynamics of chaotic time series is not masked by additive measurement noise with different noise variance unlike the research reported in Rosenstein *et al.* [56]. But here the dynamics of chaotic attractor is determined by random fluctuation in the key system parameters (*δ*). As discussed above, the system has a similar framework like a noise induced chaotic system [52], [53], [54], [55]. Therefore the exploration reported here is markedly different from that reported with additive noise [51] in various known family of chaotic systems. In addition, the effect of fractional dynamics on such random parameter switched chaotic systems and their corresponding Lyapunov exponents have also been investigated here.

The attractor's dynamics is normally reconstructed from a single time-series representation of the measured state $x_i$, using the time delay embedding method. If $x(\tau)$ be the evolution of the state from some initial condition $x(0)$, then the largest Lyapunov exponent is calculated as (11).

$$\lambda = \lim_{\substack{\tau \to \infty \\ \varepsilon \to 0}} \frac{1}{\tau} \ln\left(\frac{|x(\tau) - x_\varepsilon(\tau)|}{\varepsilon}\right), \quad |x(0) - x_\varepsilon(0)| = \varepsilon \tag{11}$$

Now for varying relative time ($\tau$) the Lyapunov exponents are calculated as (12).

$$S(\tau) = \frac{1}{N} \sum_{i=1}^{N} \ln\left(\frac{1}{|U_i|} \sum_{x_j \in U_i} \text{dist}(x_i, x_j, \tau)\right) \tag{12}$$

where, $U_i$ is the neighborhood of $x_i$ with diameter $\varepsilon$ and the $\text{dist}(x_i, x_j, \tau)$ is the distance between a trajectory $x_i$ and a neighbor $x_j$, after the relative time $\tau$. The plot of $\tau$ vs. $S(\tau)$ helps to determine whether the embedding dimension is selected in a proper way depending on whether the curve is more or less horizontal or smooth [57]. Gao *et al.* [58] proposed a technique using scale-dependent Lyapunov exponent to distinguish between chaos, noisy chaos and noise induced chaos where the presence of the chaos could be confirmed by observing a plateau over multiple scales. The present work takes a similar approach where the Lyapunov exponents with varying relative time ($\tau$) is reported as also shown in [57]. It is a well discussed topic to distinguish chaos from noise using Lyapunov exponents and there are significant amount of research done on this topic e.g. [59], [60], [61], [62]. Kinser [60] reported that the LLE of noise may be a very large number and other nonlinear measures may be needed. Therefore in the present scenario, the LLE, Lyapunov exponents for varying relative time $S(\tau)$ and Shannon entropy (*H*) analysis have been reported together, in order to understand the nature of the time series obtained from these new classes of randomly switched parameter nonlinear dynamical systems.

Apart from the LLE, another measure widely used in the investigation of randomness of the information and nonlinearity is the Shannon entropy. It can be easily calculated from the wavelet decomposition of the time-series $x(t)$ in several orthogonal basis ($x_i$) such that $H(x) = \sum_i H(x_i) = \sum_i x_i^2 \log(x_i^2)$ along with the convention of $H(0) = 0\log(0) = 0$. Here, the Lyapunov exponents and wavelet entropy were calculated using a four times down-sampled version of the recorded time series of the first state variable (*x*) using the time delay embedding technique reported in [56]. A 50 sec of simulation time with fixed step-size of $10^{-3}$ sec in four time down-sampled version would therefore generate 12500 samples, for the calculation of features i.e. LLE and Shannon entropy as the two potential system diagnostics. This down-sampling of the state time series were required due to the issue of memory overflow while calculating the LLE using large number of samples. In order to obtain a fair comparison of these estimates, the LLE and Shannon entropy for a white Gaussian noise and chaotic Logistic map as in (13) are also calculated as two extreme examples of pure noisy and pure chaotic dynamics respectively.

$$\text{Logistic map}: x_{n+1} = rx_n(1-x_n), r=4, x_1 \in \mathcal{U}(0,1)$$
$$\text{White Gaussian Noise}: x_n \sim \mathcal{N}(0,\sigma^2), \sigma=1 \tag{13}$$

Such a benchmarking of standard deterministic and stochastic dynamics (along with their overlap) is helpful in the intuitive understanding of the proposed randomly switched parameter chaos in commensurate FO nonlinear dynamical systems, exhibiting new class of semi-deterministic dynamics. In the two deterministic/stochastic discrete time iterators in (13) , the initial starting value is chosen from a uniform and normal distribution respectively while other parameters ($r$ for logistic map and standard deviation $\sigma$ for white noise) are kept constant. Before the calculation of LLE and entropy, the state time series or sequences from discrete iterators were standardized using the transformation $\tilde{x} = (x-\mu)/\sigma$ to ensure that the data is zero mean and unit variance to avoid any biased estimation. The calculated LLE and entropy show a clear signature of chaos for the Logistic map due to significantly large and positive LLE while the high randomness is affirmed from the large entropy value for white Gaussian noise. Also, the behavior of sequences from FO coloured noise (with $\alpha$ = 0.9) and noisy chaos ($\alpha$ = 0, 0.9) with SNR = 1 have been quantified for comparison, where the signal corresponds to the chaotic time series and noise is white or coloured of the same time length. When adding two sequences (noisy or chaotic) both of them are standardised separately and then added. We have used the Grunwald-Letnikov numerical fractional derivative (14) to generate fractional Gaussian noise (fGn) with $\alpha$ = 0.9 by passing a white noise sequence through a FO differentiator ($s^\alpha$).

$$_aD_t^\alpha f(t) = \lim_{h \to 0} \frac{1}{h^\alpha} \sum_{j=0}^{\left[\frac{t-a}{h}\right]} (-1)^j \binom{\alpha}{j} f(t-jh)$$
$$\binom{\alpha}{j} = \frac{\alpha!}{j!(\alpha-j)!} = \frac{\Gamma(\alpha+1)}{\Gamma(j+1)\Gamma(\alpha-j+1)} \tag{14}$$

For all the three subclasses of the unified chaotic systems i.e. fixed parameter FO Lorenz-Lu-Chen, the LLE decreases and entropy increases with gradual decrease in the commensurate FO $\alpha$ as shown in Table 3. The disappearance of chaos is also evident from sudden drop in the LLE (from change in second decimal place to the first decimal place) at $\alpha$ = 0.8 for the fixed parameter FO unified system. A similar behavior is expected for the random parameter case also where the chaos disappears much faster at $\alpha$ = 0.85.

These time series based estimates of LLE and entropy affirms the graphical findings in previous sections. In principle, for stable nonlinear systems i.e. damping of chaotic oscillations should exhibit a negative LLE but in most cases when the chaos stabilizes, it still show a positive LLE which should not be confused with weak chaotic behavior. Since, the estimation of LLE is attempted from the finite number of samples of the state time series and not the system structure itself (using Jacobian), the involved numerical method may not often give exactly zero or negative value but gives a LLE which is close to zero and also a sudden drop in the estimated Lyapunov exponents for varying relative time. The motivation behind applying a single state time series based LLE calculation instead of the conventional structure based LLE estimation method is that under varying structure or random switching between different ODE structures the latter technique does not hold [63], [64]. Therefore, using the observed time series of the state variables via the time delay embedding is a viable alternative to compute the LLE for these new interesting FO semi-deterministic chaotic systems. The delay embedding techniques are mostly used on experimental data when the underlying data generation process or the governing differential equations are not precisely known or there is noise in the measurement system. Although it is not very popular for simulation studies on deterministic chaos, as a structure based LLE computation is preferred without the presence of noisy fluctuations.

Next the Lyapunov exponents for varying relative time - $S(\tau)$ is explored [57] using equation (12), for the fixed parameter systems in Figure 11 and the random parameter system in Figure 12, while the relative time has been increased from $\tau = [1, 20]$. Figure 11 shows that the Lyapunov exponents have significantly positive value for α = 1 and α = 0.9 for all the three cases of Lorenz, Lu and Chen system. But with α = 0.8, the estimated Lyapunov exponents are much smaller and sometimes become negative, especially at higher value of $\tau$. In Figure 12, the Lyapunov exponents of a single realization of the random parameter unified system show that although there is a strong chaotic behaviour for α = {1, 0.95, 0.9}, but at α = 0.85 the chaotic behaviour is lost due to drop in the Lyapunov exponent values which confirms the studies reported in the previous section. These $S(\tau)$ plots along with the LLE estimates confirm the presence or disappearance of chaos in each FO system.

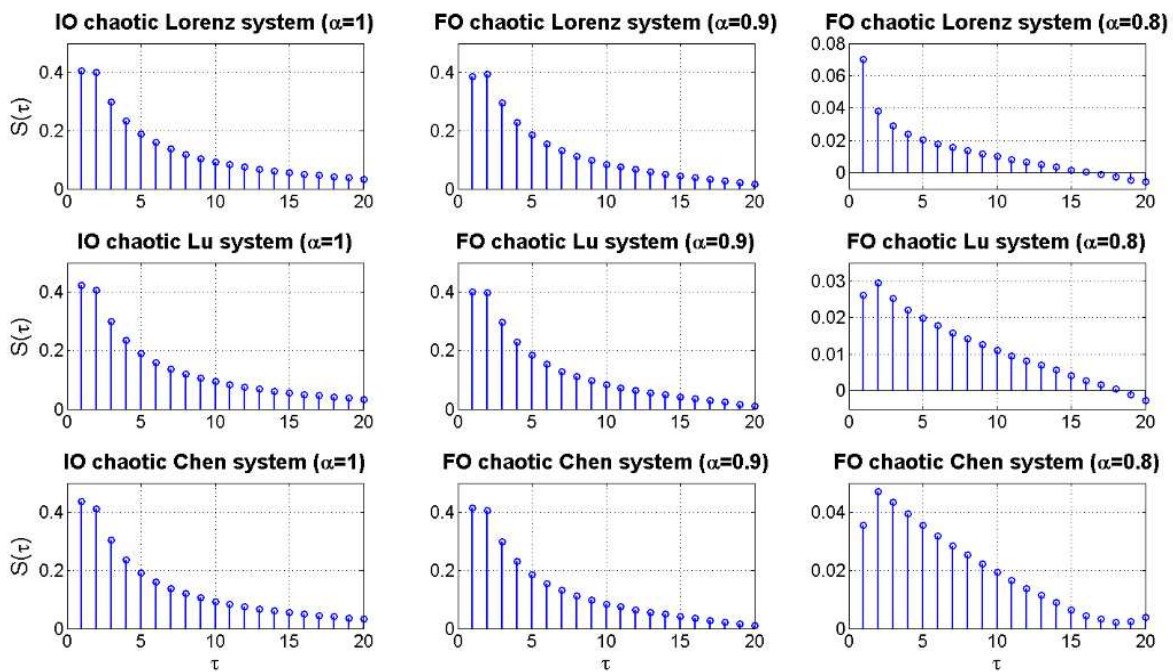

*Figure 11: Lyapunov exponents for fixed parameter chaotic systems.*

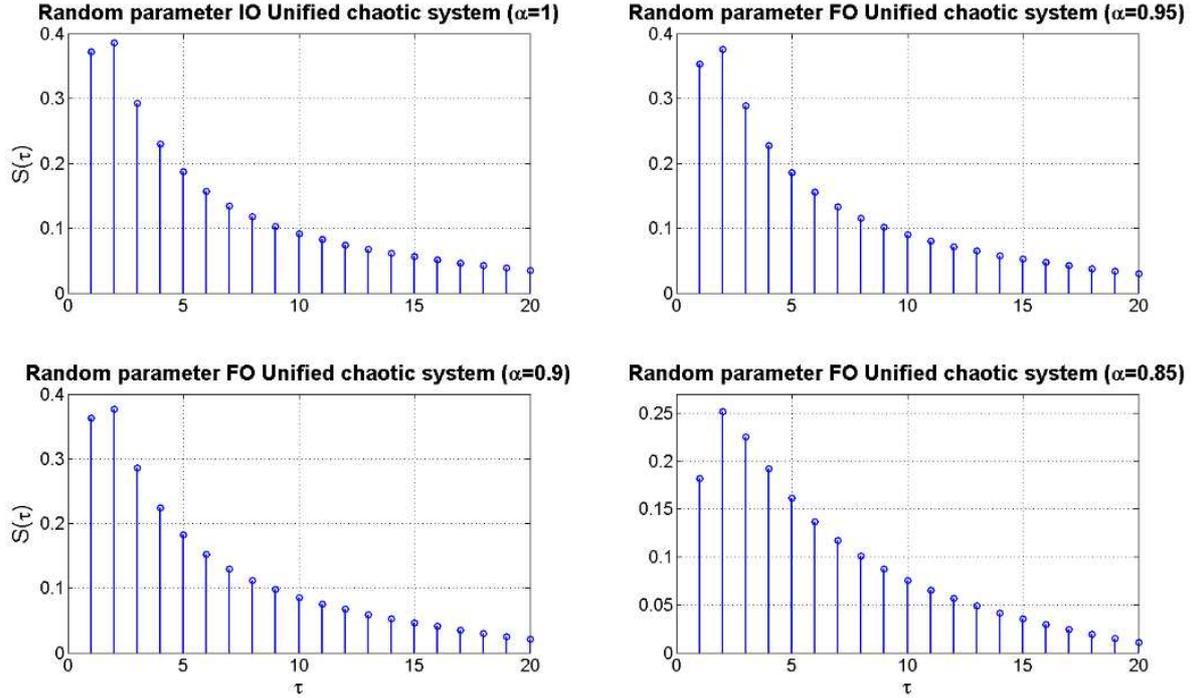

*Figure 12: Lyapunov exponents for a single realisation of random parameter chaotic systems*

*Table 3: LLE and Shannon entropy for the fixed parameter chaotic systems based on phase space reconstruction (PSR) of the first state*

| System | LLE from PSR of state $x$ | Shannon Entropy |
|---|---|---|
| IO chaotic Lorenz system ($α = 1$) | 0.1122 | -26534.90 |
| FO chaotic Lorenz system ($α = 0.9$) | 0.1105 | -23337.01 |
| FO chaotic Lorenz system ($α = 0.8$) | 0.0115 | -295631.19 |
| IO chaotic Lu system ($α = 1$) | 0.1142 | -31135.96 |
| FO chaotic Lu system ($α = 0.9$) | 0.1165 | -23748.88 |
| FO chaotic Lu system ($α = 0.8$) | 0.0089 | -305190.45 |
| IO chaotic Chen system ($α = 1$) | 0.1157 | -29141.11 |
| FO chaotic Chen system ($α = 0.9$) | 0.1188 | -24788.91 |
| FO chaotic Chen system ($α = 0.8$) | 0.0157 | -314398.57 |

For the random parameter cases, the LLE and entropy are likely to be different for each independent realisation of the system. Therefore, we have quantified them in the scatter diagrams and the associated histograms. Here, we have reported the comparison of the characterization of chaotic systems using LLE and Shannon entropy for the two frequency domain and four time domain methods of numerically solving chaotic FODEs as discussed in section 2. We have also explored the variation in system diagnostics like LLE and Shannon entropy for 100 Monte Carlo runs of the random parameter switching of the unified system. Scatter diagram of these diagnostics clearly show that the random parameter switching destroys the chaotic nature much earlier for FO unified system family. Similar comparisons are also done with these quantitative estimates for the discrete time chaotic and noisy iterators like fractional Gaussian ($1/f^\alpha$) noise, Logistic map with random initial value ($x_1 \in \mathcal{U}(0,1)$) and noise corrupted Logistic map (corrupted with white and $1/f^\alpha$ noise with SNR = 1).

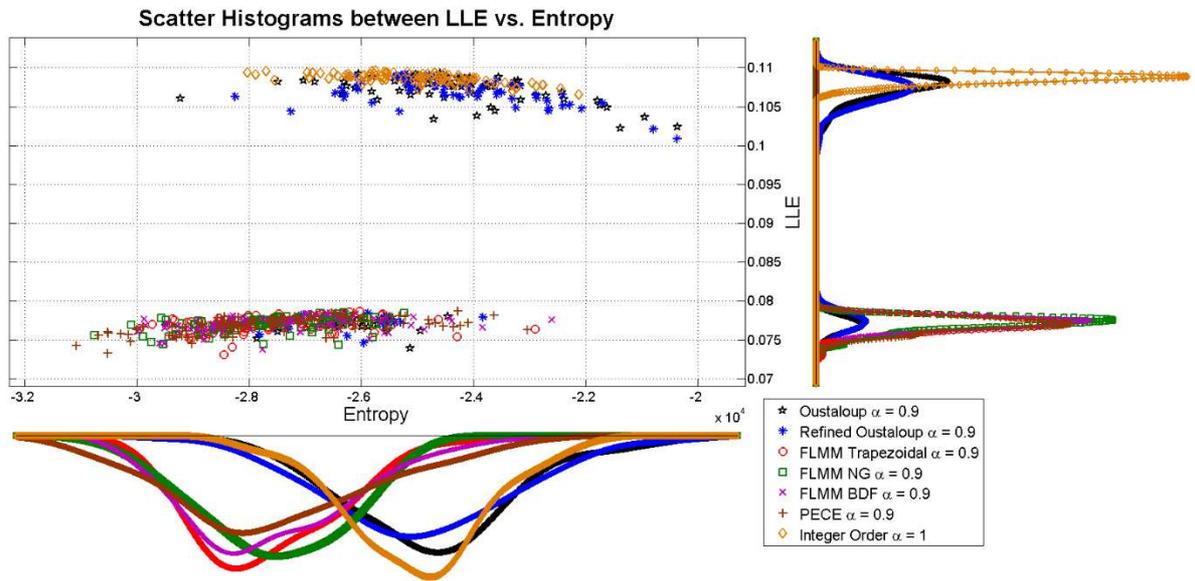

*Figure 13: Scatter histogram between LLE and Entropy for 1000 Monte Carlo runs of FO random parameter switched chaotic system with α = 0.9*

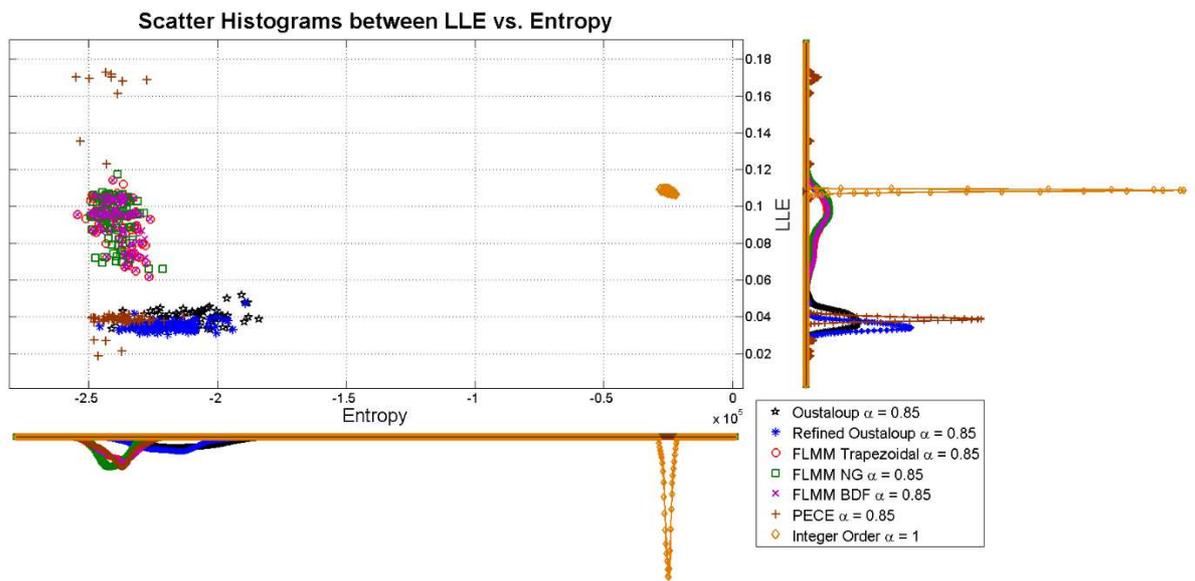

*Figure 14: Scatter histogram between LLE and Entropy for 1000 Monte Carlo runs of FO random parameter switched chaotic system with α = 0.85*

The two dimensional scatter diagrams shown in Figure 13-Figure 16 are especially useful in understanding how the quantitative measure of chaotic behaviour (LLE) and randomness (entropy) change with the decrease in the commensurate FO for the random parameter systems. These scatter diagrams also explains the trade-off between the deterministic and stochastic dynamics under the same commensurate order $\alpha$ while integrated with different time or frequency domain solvers for FODEs. Robustness of a particular method can be quantified by studying the variance of these distributions along each of the features.

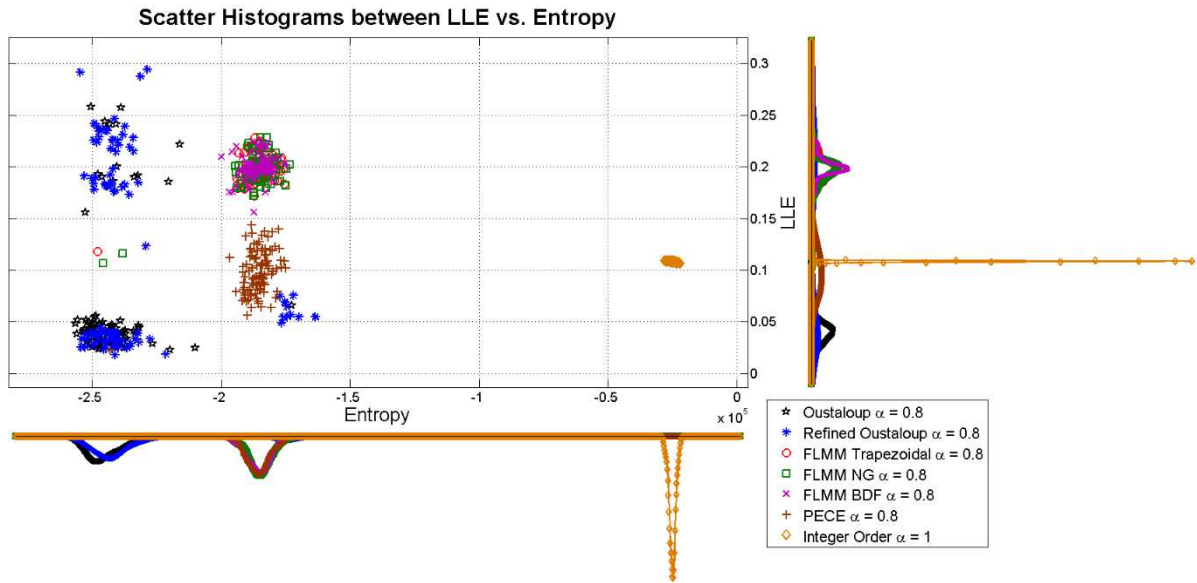

Figure 15: Scatter histogram between LLE and Entropy for 1000 Monte Carlo runs of FO random parameter switched chaotic system with $\alpha = 0.8$

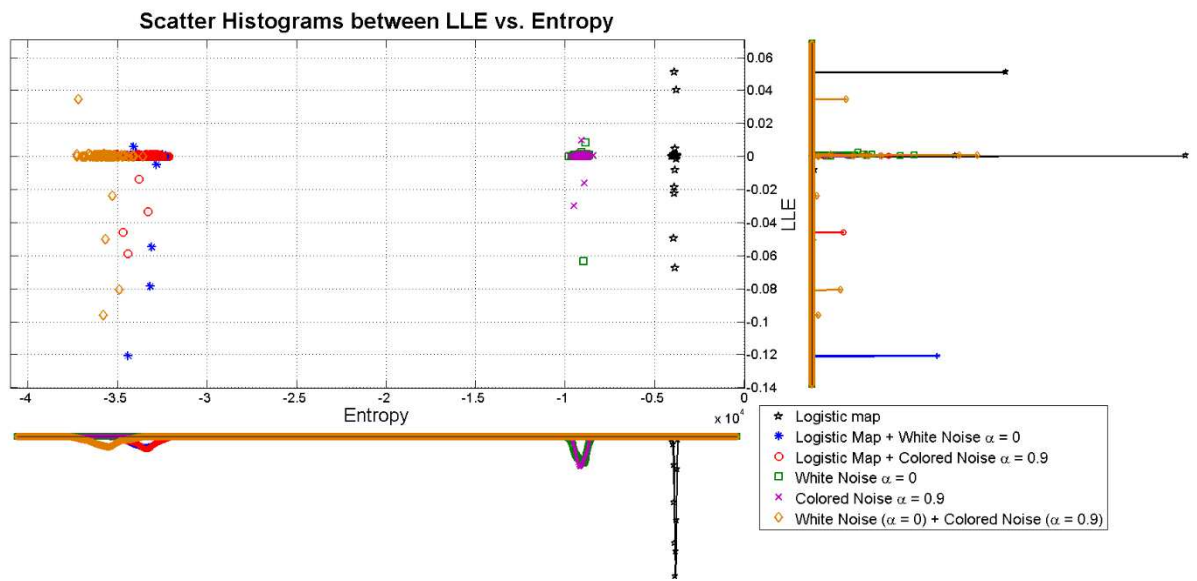

Figure 16: Scatter histogram between LLE and Entropy for 1000 Monte Carlo runs of discrete time iterators (chaotic, noisy sequences and chaos with additive noise)

A close observation in Figure 13-Figure 15 reveal that the random parameter IO chaotic system yields a robust estimation of LLE as evident from the high probability peaks. In general the FLMM algorithms also estimates the LLE with low variance compared to the other time and frequency domain methods for $\alpha = 0.9$. For further reduction in the commensurate FO the chaotic behaviour disappears and the peaks in the LLE or entropy histograms are no longer found as prominent as the IO case. It is also evident from Figure 13 that the two frequency domain method slightly over-estimates the LLE as 0.07-0.08, compared to the LLE range obtained using time domain methods 0.1-0.11 for $\alpha = 0.9$. Here the entropy does not discriminate much amongst the time and frequency domain methods and produces overlapping histograms in Figure 13.

*Table 4: Characterization of different continuous and discrete chaotic systems with noise using Shannon entropy and LLE (modes of 100 Monte Carlo runs)*

| System | Numerical Integration Solver | Shannon Entropy | LLE | Computation time (sec) |
|---|---|---|---|---|
| Logistic Map | Discrete iteration | -4020.754 | 0 | 321.812 |
| Logistic Map + White Noise $\alpha = 0$ | Discrete iteration | -35255.276 | 0 | 321.789 |
| Logistic Map + Coloured Noise $\alpha = 0.9$ | Discrete iteration | -35143.765 | 0 | 321.908 |
| White Noise $\alpha = 0$ | Discrete iteration | -9792.310 | 0 | 320.632 |
| Coloured Noise $\alpha = 0.9$ | Discrete iteration | -9731.640 | 0 | 320.501 |
| White Noise $\alpha = 0$ + Coloured Noise $\alpha = 0.9$ | Discrete iteration | -37240.687 | 0 | 322.080 |
| Integer Order $\alpha = 1$ | Runge Kutta | -28032.864 | 0.107 | 147.303 |
| | Oustaloup + Runge Kutta | -29227.048 | 0.074 | 144.554 |
| | Refined Oustaloup + Runge Kutta | -29491.429 | 0.075 | 144.429 |
| | FLMM Trapezoidal | -29955.346 | 0.073 | 158.607 |
| | FLMM NG | -30755.774 | 0.074 | 158.390 |
| | FLMM BDF | -29998.819 | 0.074 | 159.028 |
| Random parameter with $\alpha = 0.9$ | PECE | -31080.740 | 0.073 | 158.336 |
| | Oustaloup + Runge Kutta | -240975.978 | 0.033 | 20.989 |
| | Refined Oustaloup + Runge Kutta | -245596.069 | 0.030 | 21.303 |
| | FLMM Trapezoidal | -254232.479 | 0.062 | 24.571 |
| | FLMM NG | -249219.331 | 0.066 | 24.698 |
| | FLMM BDF | -254397.185 | 0.062 | 24.678 |
| Random parameter with $\alpha = 0.85$ | PECE | -254824.825 | 0.019 | 23.129 |
| | Oustaloup | -256240.080 | 0.023 | 25.291 |
| | Refined Oustaloup | -254630.812 | 0.018 | 25.272 |
| | FLMM Trapezoidal | -247888.269 | 0.118 | 26.803 |
| | FLMM NG | -245700.750 | 0.107 | 26.721 |
| | FLMM BDF | -199977.065 | 0.156 | 26.987 |
| Random parameter with $\alpha = 0.8$ | PECE | -243117.460 | 0.023 | 25.334 |

From the 100 Monte Carlo runs of the FO random parameter switched chaotic system, mode of the distributions for LLE and Shannon entropy estimates have been reported in Table 4, for benchmarking purpose with respect to the standard discrete time iterators (chaotic and noisy). It is to be noted that the LLE reported here is based on the phase space reconstruction of a single state variable and not based on the system structure as the structure itself changes at each time instance due to random parameter switching. This helps in a fair comparison of continuous time chaotic systems with the discrete time ones (e.g. Logistic map or white noise) as if the order and structure of underlying dynamical system is unknown and there is only one measurable state available to characterize the system's deterministic/stochastic behavior. For a specific choice of α, it is observed that the LLE becomes lower (indicating a weaker chaos) and entropy becomes higher (stronger stochastic dynamics or influence of noise) for the random parameter systems (in Table 4) as compared to those of the corresponding fixed parameter systems in Table 3. As an example for random parameter unified system with α = 0.9 showing chaotic phase portraits, LLE drops ~ 0.07 (using different solvers) from the fixed parameter LLE ≈ 0.11, whereas the entropy increases to H ≈ -29×$10^3$ to -31×$10^3$ for the random parameter one from the fixed parameter case where H ≈ -23×$10^3$ to -24×$10^3$. This observation is just the opposite for the same system with α = 0.8, where the chaotic oscillation is damped as the LLE increases to 0.01-0.02 due to random switching from the fixed parameter LLE ≈ 0.009-0.01. Consequently increase in entropy is observed with an order of magnitude i.e. H ≈ -19×$10^4$ to -25×$10^4$ (increased for random parameter system) and H ≈ -29×$10^4$ to -31×$10^4$ (decreased for fixed parameter system). It is to be noted that the estimates reported here are not optimally tested with different signal lengths and SNR levels for brevity, which can be pursued as a scope of future research.

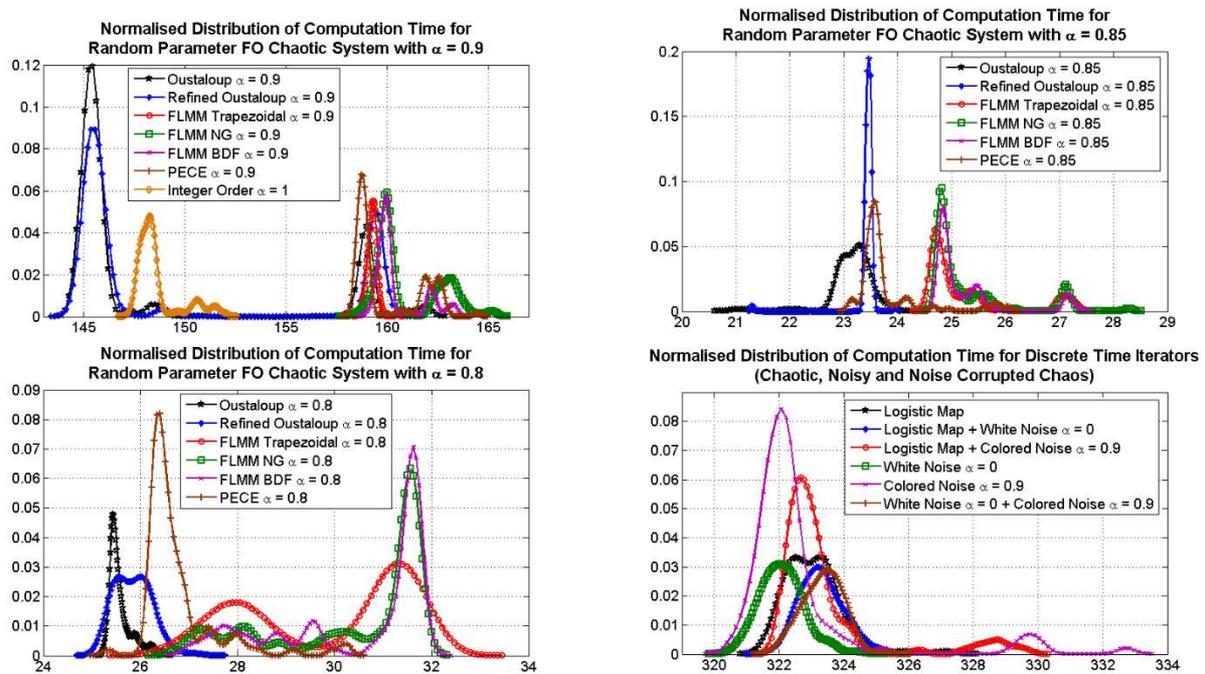

*Figure 17: Normalised distribution of computation time (in sec) for continuous (random parameter) and discrete (random initial value) chaotic systems, estimated from 100 Monte Carlo runs.*

*Table 5: Variance of the Shannon entropy and LLE estimates for 100 Monte Carlo runs of the random parameter switched unified chaotic system with α = 0.9*

| FODE Solution Technique | Variance of Shannon Entropy | Variance of LLE |
|---|---|---|
| Oustaloup + Runge Kutta | 2332101.8 | 164×$10^{-6}$ |
| Refined Oustaloup + Runge Kutta | 2646576.5 | 183.2×$10^{-6}$ |
| FLMM Trapezoidal | 1546735.2 | 0.9×$10^{-6}$ |
| FLMM NG | *1408110.1* | 0.9×$10^{-6}$ |
| FLMM BDF | 2051054.9 | *0.8×$10^{-6}$* |
| PECE | 2977598.9 | 1×$10^{-6}$ |

Histograms of the run times have also been reported in Figure 17 since the LLE computation automatically selects the optimum embedding dimension using the algorithm in Rosenstein *et al.* [56]. Therefore there is a clear difference between the LLE computation time as well, for various classes of chaotic systems and noise. It is evident from Figure 17 that the mode of the distribution for computation time of LLE is much less (within 20-35 sec) for the random parameter chaotic system with $α = 0.85$ and $α = 0.85$, whereas it is significantly higher for $α = 0.9$ (140-170 sec) as the latter shows chaotic behavior. Therefore it takes more time to find the optimum embedding dimension to be used for the numerical computation of LLE. The run times are much higher for discrete time iterators (320-325 sec) compared to the continuous time dynamical systems.

A closer look at Table 4 reveals that in most of the noisy discrete time cases, the LLE is found to be zero. The LLE estimates have low variance when the chaotic behavior is clearly visible in the phase portraits i.e. with α = 0.9. With further reduced commensurate FO (when the chaotic behavior is no longer visible in the phase portraits), the frequency domain methods and the time domain PECE method shows similar result with a significant reduction in LLE, whereas the FLMM methods still show slightly higher LLE values. In order to compare the performance of the frequency and time domain solvers for FODEs, the variance of the LLE and entropy estimates have been reported in Table 5, using the 100 Monte Carlo runs of the random parameter chaotic system with $α = 0.9$. Table 5 clearly shows that the FLMM BDF yields minimum variance in LLE estimation and FLMM NG formula yields minimum variance in the entropy estimation, In general, the time domain methods clearly scores better over the frequency domain methods, for a robust (minimum variance) LLE computation.

## 5. Discussion

### *5.1. Novelty of the Present Approach over Existing Theory of Noisy Chaos*

There have been many studies on discriminating noisy chaos, noise induced chaos and FO stochastic processes ($1/f^α$) or higher order colored noise e.g. in [58], [59]. It has also been shown in [51], that adding white or correlated noise makes a chaotic time series impossible to be identified as chaos by standard LLE method which is known as titration of chaos by noise. In such cases, the regular smooth structure of the attractor dynamics in phase space breaks down and it no longer shows chaotic oscillations. Here we explore a different phenomenon of noise induced chaos in FO unified system where the key parameter (δ) was switched randomly along with gradual decrease in the commensurate FO of the system. In the present study, at each time instant the key parameter δ was sampled from a uniform distribution that determines which family the uniform chaotic attractor belongs to at each time instant.

Also, there are other methods to investigate deterministic chaos e.g. the 0-1 test, apart from the popular method of LLE. But the 0-1 test of chaotic dynamics was originally developed to show chaotic dynamics with large number of noise free data [65]. Investigations by Gottwald and Melbourne [66] suggests that the 0-1 method can still detect the underlying deterministic chaotic dynamics if the noise-level is sufficiently small (10%) and not correlated with the systems dynamics, which is rather restricted than the problem addressed here. In our present study, the main idea is not to investigate weak chaos masked by additive noise but to characterize the systems dynamics due to a noise like parameter switching. In such cases of noise induced FO chaos, the 0-1 test is not guaranteed to detect the underlying chaotic behavior. Therefore we restrict our study to determine the domination of the deterministic and stochastic dynamics using the two popular diagnostic methods – Lyapunov exponents and Shannon entropy only. Also, further investigation is needed to understand why the chaos disappears with decrease in the commensurate FO ($\alpha$) of the system. Almost in all FO chaotic dynamical systems, the chaos is found for a certain range of parameters and not under all parameter settings. Previous investigations of FO chaos mostly reported similar approach of the existence of chaos by simulation study, under various parameter settings. The motivation of the present paper is to investigate the range of system parameters for which the chaos persists in a uniformly distributed noise induced FO unified chaotic systems, although the physical interpretation of disappearance of chaos especially at low $\alpha$ is still an open question. Also, the distribution of the key parameter ($\delta$) is chosen from a uniform distribution rather than a Gaussian or other type of distributions. This ensures almost equal number of samples drawn from all different parts of the distribution and also strictly enforces the key parameter to lie within zero to one, signifying switching amongst Lorenz, Chen and Lu families as per equation (2).

Also, different dynamical characteristics, in terms of sustained or damped excursion of states have been observed here while using different values of the commensurate FO. This raises a question whether the suppression of chaos is a generic and consistent characteristic for large scale nonlinear dynamical systems in the presence of noise and random parameter switching. Molgedey *et al.* [50] have shown that noise has different effects on the chaotic dynamics depending on the dimensionality of the system. For low dimensional systems, noise favors chaos, while in high dimensional systems noise impairs the flow of information and inhibits chaos. In this paper, the consideration of the fractional dynamics in the state equations makes the system infinite dimensional (which is approximated by a very high dimensional constant phase filtering). The aim here is to investigate whether the suppression of chaos due to noise is a generic phenomenon among all the members of the family of FO attractors (with different values of $\alpha$). Previous literature like [50], [51] addresses the stabilizability and detectability of chaos in the presence of varying degrees of additive noise which is different from random parameter switching within the three families of attractors. The reported simulation examples can be viewed as a first study for this specific type of phenomena and a more in depth study using analytical techniques like switched and hybrid dynamical systems are required in the future to lay the foundation of the underlying mathematical basis, which may be pursued as a scope for future work. One of the limitations in applying the switched chaotic dynamical systems theory [67], in the present scenario, is that the Lipschitz continuity must be maintained at each of the switching instants [68]. In other words there cannot be any kind of arbitrarily fast switching for the stabilization results to hold. There is also a consequent concept of dwell time which puts an analytical bound on the rate of switching dynamics [68]. In this paper, a noise like switching in $\delta$ at every sampling instant is considered, unlike specific time instances as reported in [67] which employ a zero order hold mechanism to satisfy the conditions of Lipschitz continuity. This might pose problems for application of the established analytical techniques in the present random parameter switched chaotic systems of integer [67] and fractional order. Future work might look at appropriate mathematical refinements to address this issue of stability analysis in the presence of arbitrarily fast switching.

## 5.2. *Physical Significance of the Observation*

Experimentally observed time domain trajectories of many real world processes as found in diverse disciplines (like finance, biology) are often characterized by noise like fluctuations. Therefore modelling the dynamical characteristics through ordinary (integer or fractional order) nonlinear differential equations inherently use some kind of approximations due to the smoothness of the differential equation solutions. Most of the literature tries to capture the noisy part by incorporating additive or multiplicative random variable in the state variables of the differential equations. The present work shows how the behavior of the FO dynamical system changes under the influence of random parameter *vis-a-vis* its constant counterpart. It has been shown in the previous sections that randomly modulating the key parameter ($\delta$) results in the disappearance of chaos. This is an important observation and may have many future applications, especially with respect to chaos control and synchronization. Recently a special kind of random parameter chaotic system, known as switched chaotic system has been explored in [67]. A single global controller has also been designed analytically based on Lyapunov stability theory for the synchronization of chaos in a wide variety of scenarios. The presented results add value to the existing knowledge of such random parameter chaotic systems of IO [67] since there might not be any need to design any external controller at all as per the present design. It may be possible to selectively modulate (by simply introducing randomness in) some key parameters of a chaotic system to stabilize it.

From perspective of mathematical foundation, the results shown in this paper bridge three major branches of research in differential equations – nonlinear chaotic differential equations, FO differential equations and stochastic differential equations. Therefore this paper can serve as a building block for further study of a new type of design paradigm which leverages the strength of these three distinct disciplines in differential equation based modelling of natural processes *viz.* sensitivity to initial condition (as modelled by chaotic nonlinear differential equations), long memory process and rich higher order dynamics (as modelled by FO differential equations) and stochastic fluctuation in the state variables (as modelled by stochastic differential equations).

## 6. Conclusion

A new class of commensurate FO systems has been explored in this paper known as random parameter switched chaotic systems. The effect of random parameter switching in the key parameter ($\delta$) of unified system under different commensurate FO have been studied. Simulation results and illustrations for the unified chaotic system show that chaotic behavior disappears much early (at a higher value of commensurate FO) for random parameter switched chaotic systems while gradually decreasing the commensurate order. The corresponding fixed parameter case still shows chaotic behavior in this regime. The introduction of noise induced parameter switching in the chaotic system and its effect in the titration of chaos [51], in other class of chaotic systems may be explored in future studies. Here, we also show that the random switching of the key parameter ($\delta$) of the FO unified system could be used to damp out the oscillations. Further engineering applications of the proposed system could be directed towards stabilization or control of chaos using the theory of switched dynamical systems. In addition, the noisy sequence for $\delta$, along with its rate of switching and the commensurate FO of the system could be used in various future applications in secure communication, encryption etc. among others.